\documentclass{aa}
\usepackage{graphicx}
\usepackage{txfonts}
\usepackage[colorlinks, linkcolor=blue, urlcolor=blue,
citecolor=blue,bookmarks, bookmarksopen, bookmarksnumbered,
pdfauthor={IAU}, breaklinks=true]{hyperref}
\usepackage{natbib,twoopt}        
\bibpunct{(}{)}{;}{a}{}{,} 
\usepackage{calc}  
\usepackage[table]{xcolor}  
\usepackage{todonotes}
\let\oldtabular\tabular
\renewcommand{\tabular}{\footnotesize\oldtabular}
\usepackage[table]{xcolor}
\newcommand\T{\rule{0pt}{2.6ex}}       
\newcommand\B{\rule[-1.2ex]{0pt}{0pt}} 
\newcommand{\X}[2]{\diaghead{CORRINDX}{#1}{#2}}
\usepackage{makecell}
\begin{document}

\title{OIFITS~2: the $2^{nd}$ version of the data exchange standard\\ for optical
  interferometry
\thanks{We request that comments and suggestions related to
  OIFITS be directed to the OLBIN email list. (See
  \url{http://www.jmmc.fr/olbin-forum} for
  information on how to subscribe and post to the list.)}}

\titlerunning{OIFITS\,2}
\author{Duvert, Gilles\inst{1,2}, Young, John\inst{3}, Hummel, Christian\inst{4}}
\institute{
 Univ. Grenoble Alpes, IPAG, F-38000 Grenoble, France.
  \thanks{Correspondence: {\tt Gilles.Duvert@univ-grenoble-alpes.fr}}
 \and
 CNRS, IPAG, F-38000 Grenoble, France.
  \and
  Cavendish Laboratory,
  University of Cambridge, J J Thomson Avenue, Cambridge, UK.
  \thanks{Correspondence: {\tt jsy1001@mrao.cam.ac.uk}}
  \and
  European Southern Observatory,
  Karl-Schwarzschild-Str.~2, 85748 Garching
  \thanks{Correspondence: {\tt chummel@eso.org}}
}

\date{Received; Accepted}

\abstract{ This paper describes version~2 of the Optical
  Interferometry exchange Format (OIFITS), the standard for exchanging
  calibrated data from optical (visible or infrared)
  interferometers. This IAU-endorsed standard has been in use for 10
  years at most of the past and current optical interferometer
  projects, including COAST, NPOI, IOTA, CHARA, VLTI, PTI and the Keck
  interferometer. Software is available for reading, writing and
  merging OIFITS files. This version~2 provides
  definitions of additional data tables (for example for polarisation
  measurements), addressing the needs of future interferometric
  instruments.  Also included are data columns for a more rigorous
  description of measurement errors and their correlations.  In that,
  this document is a step towards the design of a common data model
  for optical interferometry.  Finally, the main OIFITS header is
  expanded with several new keywords summarising the content to allow
  doing data base searches.  }

\keywords{methods: data analysis -- techniques: interferometric
  -- instrumentation: interferometers
  -- techniques: image processing -- virtual observatory tools}

\maketitle

\section{Introduction}
The exchange of data from optical and infrared interferometers between
researchers has been greatly simplified by adopting a common format
defined by \citet{Pauls2005} and now known as OIFITS version\,$1$.

Since then, new interferometric instruments have introduced new
capabilities such as sensitivity to polarisation, and new imaging
techniques for multi-channel data have been developed that require the
source spectrum -- both of which were not contemplated in the first
version of this format. Furthermore, as modern data archives such as
the ESO Science Archive \citep{2008ASPC..394..235D} require the presence
of keywords in the FITS headers to enable searches based on, for example,
target and instrument, such keywords need to be defined specifically
for Optical Interferometry (OI).

Therefore, a revision of the standard became desirable
\citep{2008SPIE.7013E..4HY} and improvements
have been discussed for two years on the JMMC
forum at \url{www.jmmc.fr/twiki/bin/view/Jmmc/OIFITSTwoProject}.

The proposed enhancements have been endorsed by the former IAU
Commission 54 and are described in the following text, including the
formal definition of all extension tables, revised or not, for
completeness. The reference for the new format shall consist of the
original paper by \citet{Pauls2005} and this work.

In the following, the use of words like ``must'' and ``should'' follow
the best current practice outlined in \citet{RFC2119}.
\section{Purpose and Scope} 
\label{sect:purpose}
By defining and maintaining a common data format, we hope to encourage
the development of common software for the analysis of data from
optical interferometers, and to facilitate the exchange of data
between different research groups.

The format was initially intended to only support the storage of
calibrated, time-averaged data, and this is still its purpose
today. Fully supporting other uses, such as describing raw data,
observational history, data reduction recipes, etc\ldots will require
the development of a specific Data Model in the framework of the
Virtual Observatory (VO, see \url{http://www.ivoa.net}), which is
beyond the scope of this paper.

However, even since version\,$1$, the information stored in the OIFITS
format was not entirely devoid of information about the instrument and
interferometric array used. Practice, in particular for the AMBER
instrument of VLTI \citep{2007A&A...464....1P}, has shown that OIFITS
is adequate for storing semi-calibrated observations. Those are time
series of instantaneous interferometric observables where instrumental
bias has been removed and where error bars account for instrumental
noise, known as level one (L1) data in the VO jargon.

Although the format presented here defines new keywords and tables
that allow a more detailed description of the measured data, they are
intended only for storing fully calibrated data: time averaged data,
with all atmospheric and instrumental biases removed, known as L2 data in the
VO jargon. Astronomers should use only
fully calibrated OIFITS data to make inferences about the
intrinsic brightness distribution of the observed target.

Interferometers are generally polarizing devices, but useful polarized
optical interferometric observables have not been defined yet. However, we
felt compelled to introduce some support for polarisation
information. Without loss of generality, considering a
number (hereafter \texttt{NPOL}) of distinct polarisation states for
the observations leads to measurement of \texttt{NPOL} times all the
interferometric observables. For example, if a Wollaston prism is used
to make measurements on two orthogonal polarisation axes, we have
\texttt{NPOL=2}. 

In this document, we aim to keep the revised OIFITS tables compatible
with their version\,$1$. The only supplementary requirement for
polarisation handling is that a distinct \texttt{INSNAME} must be
assigned to each of the \texttt{NPOL} polarisations. A new optional
OI\_INSPOL table is introduced.  This table serves two purposes:
\begin{itemize}
\item It stores the polarisation transfer function of each beam for
  each \texttt{INSNAME} in the general form of Jones matrices (see
  Sect.~\ref{sect:inspolar}).
\item It serves as a hub for defining a relationship between all
  the polarized \texttt{INSNAME}s. The intention is that, if this
  table is present, the data referenced by OI\_INSPOL (via
  \texttt{INSNAME}) shall be kept together during merging, copying or
  splitting of OIFITS files (unless the intention is to select a
  subset of the polarisations).
\end{itemize}

OIFITS version\,$2$ provides an optional table to store the correlations
between interferometric observables, whether in wavelength, baseline,
or time. By analogy with the \texttt{INSNAME} keyword, we introduce a
\texttt{CORRNAME} keyword that uniquely references a table, OI\_CORR,
containing the sparse matrix representation of the correlations
between selected observables of the dataset.

Finally, a drawback of OIFITS version\,$1$ is also tackled here, namely
the fact that the OI\_VIS table could contain several kinds of data,
which were not distinguished in the earlier format: correlated flux,
differential (chromatic) visibilities and phases, or simply linear
visibilities.

A reference implementation of this version of OIFITS has been
developed by one of us (J. Young) in C, available at
\url{https://github.com/jsy1001/oifitslib}. We anticipate that
implementations in other languages (IDL, Python, Java\ldots) will be
available in the near future.
\section{Definitions and Assumptions}
\label{sect:defn}
In this section we only present definitions related to new quantities
stored by the format, and refer the reader to \citet{Pauls2005} for
all other definitions and assumptions. In this text the word ``error''
refers to the square root of a (temporal, spatial, theoretical\ldots)
variance.
\subsection{Differential Chromatic Visibility}
Differential chromatic visibility (in Table~\ref{tbl:oivis}) is
defined here as the ratio of visibility amplitudes
\begin{equation}
\label{eq:diffvis}
{V_{\mathrm{diff}}(u,v,\lambda) = \frac{|V(u,v,\lambda)|}{|V_{\mathrm{ref}}(u,v,\lambda)|}}~,
\end{equation}
where $|V_{\mathrm{ref}}(u,v,\lambda)|$ is referred to as the
\emph{reference visibility} which is computed as a function of one or
more \emph{reference channels}. Thus, we consider here normalisation only
to a reference derived from the data values themselves, rather than
from model predictions. Such normalisation can
be useful in cases where absolute calibration cannot be achieved, but
where spectral features relative to a continuum level are not
affected.  One therefore needs to know, for each wavelength channel,
which other channel had been used as reference.  In case multiple
channels have been used as the reference, the order of a wave-number
polynomial function used to interpolate the reference visibility for
the channel to be normalized shall be specified. The default order is
zero, equivalent to a simple average
\begin{equation}
\label{eq:average}
{|V_{\mathrm{ref}}(u,v,\lambda)| =
<|V(u,v,\lambda)|>_{\Delta\lambda}}~.
\end{equation}
\begin{table}[b]
\caption{Keywords in OIFITS binary extension table headers.}
\label{tbl:tablekw}
\begin{tabular}[c]{l c l l}
\hline\hline
\T \hfill Keyword \hfill & Req. &\hfill Reason \hfill&\hfill Example\hfill \B\\ 
\hline
\T TTYPEn$^{(1)}$& yes & name of the column  & T3PHIERR\\
TUNITn$^{(1)}$ & yes$^{(2)}$ & unit of datum & deg \\
TDIMn$^{(1)}$ & no & dimensionality of the data & (3,4) \\
DATASUM	 & no & 					&	\\
CHECKSUM  & no & 					&	\B\\
\hline\hline
\end{tabular}
\par\T{\textbf{Notes.} $^{(1)}$~``n'' stands for the column number. $^{(2)}$~may be absent for unitless columns such as indexes.}
\end{table}
\subsection{Differential Chromatic Phase}
Differential chromatic phase \citep{2006EAS....22..379M} (see
Table~\ref{tbl:oivis}), is a phase difference between two
wavelengths. An additional calibration step often applied to
differential chromatic phase is to remove a wave-number slope or a
higher-order polynomial
from the observed phase $\phi(u,v,\lambda)=\arg{V(u,v,\lambda)}$\,:
\begin{equation}
\label{eq:diffphase}
\phi_{\mathrm{diff}}(u,v,\lambda) = \phi(u,v,\lambda) -
\phi_{\mathrm{ref}}(u,v,\lambda) - \frac{\eta}{\lambda} -
\frac{\zeta}{\lambda^2} + ... ,
\end{equation}
 where $\phi_{\mathrm{ref}}(u,v,\lambda) =
\arg{V_{\mathrm{ref}}(u,v,\lambda)}$ is the phase at some adequately
chosen reference channel that can, as in Eq.~\ref{eq:diffvis}, be an average over
multiple physical wavelength channels.
\subsection{Noise model for complex visibility}
Complex visibilities are the result of a coherent average which has
removed the systematic phase noise induced by the atmosphere, either
enabled by a fringe tracker, or offline analysis of dispersed
fringes. At the level of estimates of the complex visibility from raw
frames of the interferograms, the real and imaginary parts are
uncorrelated (such as frames from an ABCD beam combiner,
see~\cite{2009A&A...498..601B}). The calibration of the averaged
product causes the noise ellipse to be elongated, as amplitude and
phase are subject to different sources of systematic error. For this
reason, version\,$1$ of the standard implemented an amplitude and phase
representation of the (average calibrated) complex
visibilities. Upcoming new instruments such as GRAVITY
\citep{2011Msngr.143...16E} will allow visibility phases to be tied to
a phase reference (nearby star), providing a data reduction chain of
complex visibilities without ever converting them to modulus of the
amplitude and phase. For that reason, we propose a new revision of the
OI\_VIS table which includes, in addition to the existing columns,
optional storage for the real and imaginary parts of complex
visibilities and their errors.
\begin{table*}[t]
  \caption{OIFITS main header keywords. Keywords in bold face are standard
    FITS keywords. The type column gives the format in C-style notation, where \%c is a single character, \%s a string,
    \%d is an integer number and \%.xf is a real value with x decimal digits. Fixed values are written textually in the type column. This applies to the other Tables in this document.
    The third column indicates whether the keyword is required or not.}
\label{tbl:metadata}
\begin{tabular}[c]{l c l l l}
\hline\hline
\T Keyword	& Type	& Req?& Description		        		& Example 	\B\\
\hline
\T {\bf SIMPLE} 	& \texttt{T} 	& yes & Standard FITS format				&		\\
{\bf BITPIX} 	& \%d 	& yes & Number of bits per data pixel			&		\\
{\bf NAXIS} 	& \%d 	& yes & Number of data axes$^{(1)}$			& 0		\\
{\bf EXTEND} 	& \texttt{T} 	& yes & Extensions may be present			&		\\
{\bf ORIGIN} 	& \%s 	& yes & Institution responsible of file creation	& ESO		\\
{\bf DATE} 	& \%s 	& yes & Date the HDU was written  			& 2015-01-13T16:12:56.57	\\
{\bf DATE-OBS} 	& \%s	& yes & Start date of observation 			& 2014-10-15T18:33:26.06	\\
CONTENT 	& \%s   & yes & Must contain only the string ``OIFITS2''$^{(2)}$			& OIFITS2\\
{\bf AUTHOR}	& \%s	& no  & As defined in FITS norm           & 	\\
DATASUM	        & \%s 	& no  & HDU datasum$^{(6)}$					&	\\
CHECKSUM	& \%s 	& no  & HDU checksum$^{(6)}$					&	\B\\
\hline
\multicolumn{5}{c}{\T Keywords set to ``MULTI'' for heterogenous content of the OIFITS (see text) \B}\\
\hline
\T {\bf TELESCOP}	& \%s 	& yes & A generic identification of the ARRAY$^{(3)}$	 	& CHARA		\\
{\bf INSTRUME}	& \%s 	& yes & A generic identification of the instrument$^{(3)}$	& VEGA		\\
{\bf OBSERVER}	& \%s 	& yes & Who acquired the data$^{(3)}$	& S.M.Eisenstein		\\
{\bf OBJECT}	& \%s 	& yes & Object Identifier$^{(3)}$		& HD123456	\\
INSMODE	        & \%s 	& yes & Instrument mode$^{(3)}$	& Low\_JHK	\\
{\bf REFERENC}	& \%s	& no  & Bibliographic reference$^{(3)}$	& 2017A\&A...597A...8D	\\
PROG\_ID 	& \%s   & no  & Program ID$^{(3)}$		& 094.A-1234	\\
PROCSOFT	& \%s 	& no  & Versioned Data Reduction Software$^{(3,4)}$ 		& pndrs	2.6	\\
OBSTECH		& \%s 	& no  & Technique of observation$^{(3,5)}$			& SCAN, ABCD-PH, PH\_REF\ldots\\
\hline
\multicolumn{5}{c}{\T Suggested supplementary keywords for ``atomic'' observations (see text) \B}\\
\hline
\T RA 		& \%f 	& no & Target Right Ascension at mean EQUINOX (deg) 			&		\\
DEC 		& \%f 	& no & Target Declination at mean EQUINOX (deg) 	&		\\
{\bf EQUINOX} 	& \%.1f & no & Standard FK5 (years)		 	& 2000.0	\\
RADECSYS	& \%s 	& no & Coordinate reference frame	 		& FK5		\\
SPECSYS 	& \%s 	& no & Reference frame for spectral coord.		& TOPOCENTR	\\
TEXPTIME 	& \%f 	& no & Maximum elapsed time for data point (s)	&		\\
MJD-OBS 	& \%.8f & no & Start of observations (MJD)	 		&		\\
MJD-END 	& \%.8f & no & End of observations (MJD)	 		&		\\
BASE\_MIN	& \%f 	& no & Minimum projected baseline length (m)		& 16.01		\\
BASE\_MAX	& \%f 	& no & Maximum projected baseline length (m)		& 100.12		\\
WAVELMIN	& \%f 	& no & Minimum wavelength (nm)				& 1200		\\
WAVELMAX	& \%f 	& no & Maximum wavelength (nm)			        & 1600		\\
NUM\_CHAN	& \%d 	& no & Total number of spectral channels 		& 32		\\
SPEC\_RES	& \%f	& no & Reference spectral resolution ($\lambda/{\Delta_\lambda}$)		        & 38323.56		\\
VIS2ERR		& \%f 	& no & Representative $V^2$ error (\%)		        & 5.3		\\
VISPHERR	& \%f 	& no & Representative Diff. Vis. Phase error (deg)	& 1.08		\\
T3PHIERR	& \%f 	& no & Representative Closure Phase error (deg)		& 32.65		\B\\
\hline\hline
\end{tabular}
{\par\T\small \textbf{Notes.} $^{(1)}$~Should be ``0'' unless a FITS image or cube is present after the main header. $^{(2)}$~Mandatory. Identifies the FITS file as an OIFITS version\,$2$ file. $^{(3)}$~These are bookeeping keywords. Some of them can be inferred from the headers of the tables contained in the OIFITS file. If the file is of ``atomic'' type (see text), all these keyords must be filled up by the relevant value. If the OIFITS acts as a container for heterogenous data and thus more than one value of the keyword would be needed, use ``\texttt{MULTI}''. $^{(4)}$~Data Reduction Software (DRS) with version identification. $^{(5)}$~Identifiers for \texttt{OBSTECH} are not standardized here. $^{(6)}$~see Sect.~\ref{sect:fitsstruct}. }
\end{table*}
\section{OIFITS File Structure}
\label{sect:fitsstruct}
OIFITS files must comply with the FITS standard \citep{FITS}, where shall be found acronyms and
  symbols not defined in this document. OIFITS files use binary table extensions following the
primary header-data unit. Data types for each column are defined by the
value of the \texttt{TFORM}n keywords in each extension headers. 
This version introduces the use of the data type J (32 bit integers).

In addition to the mandatory FITS binary extension keywords defined in
\citet{FITS}, the keywords described in Table~\ref{tbl:tablekw} must
be present in the table headers. We recommend the use of
\texttt{DATASUM} and \texttt{CHECKSUM} in the main header and all
table headers, as described in
\url{http://fits.gsfc.nasa.gov/registry/checksum/checksum.pdf}. \texttt{TUNIT}n
values are mandatory for all columns except those representing
unitless values (such as \texttt{STA\_INDEX}), which are represented by the FITS null string (all
spaces). Any other keyword is expected to be ignored by OIFITS
readers. The FITS format defines a null value for each supported data
type. The OIFITS format uses these type-dependent null values to
identify individual data as being unmeasured, invalid, etc\ldots and
this document refers to such values as \texttt{NULL}.
\begin{table}[bp]
\caption{OI\_TARGET (revision~$2$)}
\label{tbl:oitarget}
\begin{tabular}[c]{l c l}
\hline\hline
\T \hfill Label \hfill &\hfill Data Type\hfill&\hfill Description\hfill \B\\ 
\hline
\multicolumn{3}{c}{\T Keywords \B}\\
\hline
\T OI\_REVN  & \texttt{2} & Revision number:~$2$\B\\
\hline
\multicolumn{3}{c}{\T Column Headings (one row for each source)\B}\\
\hline
\T 
TARGET\_ID & I (1) & Index number. Must be $\ge 1$\\
TARGET     & A (32) & Target name$^{(2)}$\\
RAEP0      & D (1) & R.A. at mean EQUINOX (deg) \\
DECEP0     & D (1) & DEC. at mean EQUINOX (deg) \\
EQUINOX    & E (1) & Equinox$^{(1)}$\\
RA\_ERR    & D (1) & Error in R.A. (deg)\\
DEC\_ERR   & D (1) & Error in Decl. (deg)\\
SYSVEL     & D (1) & Systemic radial velocity (m\,s$^{-1}$)\\
VELTYP     & A (8) & Reference for radial velocity: \\
& &(``LSR'', ``GEOCENTR'', etc.)\\
VELDEF     & A (8) & Definition of radial velocity:\\
& & (``OPTICAL'', ``RADIO'')\\
PMRA       & D (1) & Proper motion in R.A. (deg\,yr$^{-1}$)\\
PMDEC      & D (1) & Proper motion in Decl. (deg\,yr$^{-1}$)\\
PMRA\_ERR  & D (1)  & Error of proper motion in R.A. (deg\,yr$^{-1}$)\\
PMDEC\_ERR & D (1)  & Error of proper motion in Decl. (deg\,yr$^{-1}$)\\
PARALLAX   & E (1)  & Parallax (deg)\\
PARA\_ERR  & E (1)  & Error in parallax (deg)\\
SPECTYP    & A (32) & Spectral type$^{(2)}$\\
CATEGORY   & A (3) & (optional) ``CAL'' or ``SCI''\B\\
\hline\hline 
\end{tabular}
{\small\T\textbf{Notes.} $^{(1)}$~The use of EQUINOX = $2000.0$ is recommended. $^{(2)}$~strings of length lesser than 32 are allowed for backwards compatibility, but should trigger a warning on reading the file.}
\end{table}
\subsection{Primary Header}
While the primary header was minimal in OIFITS version\,$1$, we have
added in Table~\ref{tbl:metadata} three small sets of keywords. They
add ancillary information on, for example, the provenance of the OIFITS file
itself, (\texttt{ORIGIN}, \texttt{AUTHOR}), of the data
(\texttt{TELESCOP}, \texttt{INSTRUME}, \texttt{REFERENC},
\texttt{PROG\_ID}), on the observational method used
(\texttt{INSMODE}, \texttt{PROCSOFT}, \texttt{OBSTECH}).

It is customary now to add in a FITS primary header other keywords
enabling quick searches for observations matching certain criteria in
a database or archive of OIFITS files, such as the JMMC's
  optical interferometry database, at
  \url{http://oidb.jmmc.fr/index.html}. For example, the information
about minimum and maximum baseline lengths used in the OIFITS file may
be important for selecting only data resolving a target or providing a
good $uv$-coverage for imaging. As the majority of instruments now
provide data for multiple wavelength channels, the wavelength range
recorded is a useful search parameter, as is the spectral
resolution. Finally, the maximum elapsed time for a data point can be
important information when searching for data on a wide binary where
fringe smearing is a concern. These values are easily computable
on-the-fly from the tables contained in the OIFITS file, since the
amount of (reduced) data in an OIFITS file is still small.

For OIFITS writers needing to add such ancillary information in the
main header, we suggest using the additional keywords listed in
Table~\ref{tbl:metadata}.  These keywords are intended primarily for
``atomic'' OIFITS (one target, one instrument and instrument mode, one
observing date, etc\ldots) where there is no need to describe multiple
(keyword value \texttt{MULTI}) observations.  It is out of the scope of
this document to further normalize the main header keywords for cases
where the OIFITS file is a container of heterogenous data, those
properties being more related to the problems of database management
than to a physical description of interferometric
observables. Nevertheless, see
\url{http://www.eso.org/sci/observing/phase3/p3sdpstd.pdf} for a candidate
standardisation of such ancillary keywords.
\subsection{Table content}
A valid OI exchange-format FITS file must contain one (and
only one) OI\_TARGET table, one or more OI\_ARRAY tables, one or more
OI\_WAVELENGTH tables, plus any number of the data tables: OI\_VIS,
OI\_VIS2, or OI\_T3. Each data table must refer to an
OI\_WAVELENGTH and OI\_ARRAY table that are present in the file. The
requirement for the presence of an OI\_ARRAY table was not part of
version\,$1$ of the format. All other new tables are optional.  The
revision number of the revised tables is equal to two, while it is
equal to one for the new tables. Any future changes will require
increments in the revision numbers of the altered tables.

If multiple tables of the same \texttt{EXTNAME} are present, each of
them must have a unique value of the \texttt{EXTVER} keyword, as
mandated by the FITS standard.

The tables can appear in any order. Other header-data units may appear
in the file, provided their \texttt{EXTNAME}s do not begin with
``OI\_''.  Each instance of a table defined here must contain
all the listed columns and header keywords, in no particular order.

Any of the tables may have extra keywords or columns beyond those defined
in the standard. It would facilitate the addition of new keywords and
columns in future releases of the standard if the non-standard keywords
and column names were given a particular prefix, such as ``NS\_'', to avoid
conflicts.
\section{Target and instrument information tables}
\subsection{OI\_TARGET (revision~2)}
\label{sect:target}
Keywords for the OI\_TARGET table are listed in
Table~\ref{tbl:oitarget}.  The only differences from revision\,$1$ are
the optional additional column giving the category assigned to a
target for the purpose of calibrating the data, and the recommendation
to use \texttt{EQUINOX}\,$2000$
for coordinates.  

It is standard practice to interleave observations
of science targets (\texttt{SCI} category) with observations of
targets of known angular size (\texttt{CAL} category), to allow to
calibrate the interferometric observables for the science objects.
The optional column \texttt{CATEGORY} has been introduced for
convenience as a reminder of the initial intent for observing each
target. Its presence does not imply any particular data reduction
level for the data stored in the other tables.
\begin{table}[tp]
\caption{OI\_ARRAY (revision~$2$)}
\label{tbl:oiarray}
\begin{tabular}[c]{l c l}
\hline\hline
\T \hfill Label \hfill &\hfill Data Type\hfill&\hfill Description\hfill \B\\ 
\hline
\multicolumn{3}{c}{\T Keywords \B}\\
\hline
\T 
OI\_REVN & \texttt{2} & Revision number:~$2$\\
ARRNAME  & \%s & Array name, for cross-referencing\\
FRAME    & \%s & Coordinate frame\\
ARRAYX   & \%f & \\
ARRAYY   & \%f & Array center coordinates (m)\\
ARRAYZ   & \%f &\B\\
\hline
\multicolumn{3}{c}{\T Column Headings (one row for each array element)\B}\\
\hline
\T 
TEL\_NAME  & A (16) & Telescope name\\
STA\_NAME  & A (16) & Station name\\
STA\_INDEX & I (1)  & Station number. Must be $\ge 1$\\
DIAMETER   & E (1)  & Element diameter (m)\\
STAXYZ     & D (3)  & Station coordinates w.r.t. array center (m)\\
FOV	   & D (1)  & Photometric field of view (arc\,sec)  \\
FOVTYPE	   & A (6)  & Model for FOV: ``FWHM'' or ``RADIUS''  \B\\
\hline\hline 
\end{tabular}
\end{table}
\subsection{OI\_ARRAY (revision~2)}
\label{sect:oiarray}
Keywords for the OI\_ARRAY table are listed in
Table~\ref{tbl:oiarray}.  As defined, the OI\_ARRAY tables are mostly
aimed at ground-based interferometry with separated telescopes but
could be used for other cases depending on the value of the
\texttt{FRAME} keyword. Each OI\_ARRAY table in a file must
have a unique value for \texttt{ARRNAME}. The stations involved in any
of the interferometric observables stored in the other tables of a
single OIFITS are identified via the \texttt{STA\_INDEX} column. To be
valid, an OIFITS file shall be written so that every combination of
\texttt{ARRNAME} and \texttt{STA\_INDEX} must refer to a unique
OI\_ARRAY entry that is present in the file.

Several interferometric observables stored in the OIFITS format are
expressed as a ratio of a correlated flux to an incoherent flux.  The
area on the sky contributing to the latter is the photometric field of
view (PFOV), and would typically be the PSF of a telescope in the
array, the acceptance cone of an optical fiber, the diffraction
pattern of a pinhole spatial filter\ldots  Therefore, we have added two
columns to the OI\_ARRAY table, \texttt{FOV}, which gives the radius
of PFOV and \texttt{FOVTYPE}, which specifies whether the PFOV radial
profile is Gaussian or rectangular.

The interferometric field of view (IFOV), over which photons received
contribute to the correlated flux, is smaller than the PFOV,
especially for co-axial Michelson beam combiners.  In this case, the
IFOV can be computed for each $(u,v)$ data point with the information
from the OI\_WAVELENGTH table, since in this
case $\mathrm{IFOV}=\frac{\lambda^2}{B\times\Delta_\lambda}$, where
$\lambda$ is the wavelength, $\Delta_\lambda$ the spectral width and
$B$ the projected baseline length.  The IFOV is also limited by the
integration time, which, if too long, will cause fringe smearing for
image pixels far away from the phase center.
%
\subsubsection*{Coordinate frame}
If the \texttt{FRAME} keyword has the value \texttt{GEOCENTRIC},
then the coordinates are given in an earth-centered, earth-fixed,
Cartesian reference frame. The origin of the coordinates is the
earth's center of mass. The $z$ axis is parallel to the direction of
the conventional origin for polar motion. The $x$ axis is parallel to
the direction of the intersection of the Greenwich meridian with the
mean astronomical equator. The $y$ axis completes the right-handed,
orthogonal coordinate system.

We add an alternative value for \texttt{FRAME}, namely
\texttt{SKY}, for data from sparse aperture masks, where the $x$
axis points East, the $y$ axis North and the $z$ axis completes the
right-handed coordinate frame (the aperture mask rotates with the
sky). In this case, the \texttt{ARRAYX}, \texttt{ARRAYY}, and
\texttt{ARRAYZ} coordinates must be zero.
\subsection{OI\_WAVELENGTH (revision~2)}
\label{sect:wavelength}
Keywords for the OI\_WAVELENGTH table are listed in
Table~\ref{tbl:oiwavelength}.  This table was introduced in the first
version of the format to store the effective wavelength and bandwidth
for each spectral channel of an instrument (identified by the
\texttt{INSNAME} keyword). The number of spectral channels in the
instrument is the number of rows in this table, that is, the
\texttt{NAXIS2} value, referred to as \texttt{NWAVE} in this document.

In this revision, we introduce the possibility of storing purely
monochromatic values (\texttt{EFF\_BAND}=$0$) in OI\_WAVELENGTH, since
such a table can be referenced, via the \texttt{INSNAME} keyword, by a
new table, OI\_FLUX, which may contain monochromatic spectra (see
Sect.~\ref{sect:flux}).
\subsubsection*{Name of detector}
Each OI\_WAVELENGTH table in a file must have a unique value for
\texttt{INSNAME}. Particular attention to this point is recommended when
aggregating OIFITS files which may have similar \texttt{INSNAME} values but
different channel wavebands.
\subsubsection*{Wavelengths}
Each OI\_WAVELENGTH table describes either the spectral response of
detector(s) with a number of spectral channels (traditional use
defined in version\,$1$ of the format), or the wavelengths of a
monochromatic spectrum.  In the latter case, \texttt{EFF\_WAVE}
contains the spectrum's wavelengths and the \texttt{EFF\_BAND} values
are all zeros.  In the former case, each table gives the observation
wavelengths for one or more of the data tables (OI\_VIS, OI\_VIS2,
OI\_T3), and will often correspond to a single physical detector.

The \texttt{EFF\_WAVE} column shall contain the best available estimate of the
effective wavelength of each spectral channel, and the \texttt{EFF\_BAND}
column shall contain the best available estimate of the effective
half-power bandwidth. These estimates should include the effect of the
earth's atmosphere, but not the spectrum of the target. (The effect of
the target spectrum should be taken into account as part of any
model-fitting or mapping process, in other words, the target spectrum is part of the
model.)
\begin{table}[b]
\caption{OI\_WAVELENGTH (revision~$2$)}
\label{tbl:oiwavelength}
\begin{tabular}[c]{l c p{0.5\columnwidth}}
\hline\hline
\T \hfill Label \hfill &\hfill Data Type\hfill&\hfill Description\hfill \B\\ 
\hline
\multicolumn{3}{c}{\T Keywords \B}\\
\hline
\T 
        OI\_REVN & \texttt{2} & {\tiny Revision number:~$2$}\\
        INSNAME  & \%s & {\tiny Name of detector, for cross-referencing}\B\\
\hline
\multicolumn{3}{c}{\T Column Headings (one row for each wavelength channel)\B}\\
\hline
\T 
        EFF\_WAVE & E (1) & {\tiny Effective wavelength of channel (m)}\\
        EFF\_BAND & E (1) & {\tiny Effective bandpass of channel (m)}\B\\
\hline\hline 
\end{tabular}
\end{table}
\section{Data Tables}
The tables containing observed data are the OI\_VIS2, OI\_VIS,
OI\_T3 and OI\_FLUX tables.
\subsection{Properties shared by all data tables}
\subsubsection*{Start date of observations}
This shall be a UTC date in the format \texttt{YYYY-MM-DD}, for example ``\texttt{1997-07-28}''.  In OIFITS version\,$2$ only \texttt{MJD} and
\texttt{DATE-OBS} must be used to express time.  The \texttt{TIME}
column is retained only for backwards compatibility and must contain
zeros
.  The value in the \texttt{MJD} column shall be the
mean UTC time of the measurement expressed as a modified Julian Day.
\subsubsection*{Integration time}
The exchange format will normally be used for interchange of
time-averaged data. The ``integration time'' \texttt{INT\_TIME} is
therefore the length of time over which the data were averaged to
yield the given data point. In order that this information is useful
when computing the effect of fringe smearing in the case of imaging
distant companions or large structures, \texttt{INT\_TIME} shall
correspond to the time elapsed between the first and last recorded
interferograms (frames) averaged to obtain a given data point.
\begin{table*}[t]
\caption{OI\_VIS2 (revision~$2$)}
\label{tbl:oivis2}
\begin{tabular}[c]{l c l}
\hline\hline
\T \hfill Label \hfill &\hfill Data Type\hfill&\hfill Description\hfill \B\\ 
\hline
\multicolumn{3}{c}{\T Keywords \B}\\
\hline
\T 
      OI\_REVN  & \texttt{2} & Revision number:~$2$\\
      DATE-OBS  & \%s & UTC start date of observations\\
      ARRNAME   & \%s & Identifies corresponding OI\_ARRAY\\
      INSNAME   & \%s & Identifies corresponding OI\_WAVELENGTH table\\
      CORRNAME  & \%s & (optional) Identifies corresponding OI\_CORR table\B\\
\hline
\multicolumn{3}{c}{\T Column Headings (one row for each measurement)\B}\\
\hline
\T 
      TARGET\_ID & I (1)     & Target number as index into OI\_TARGET table\\
      TIME       & D (1) & Zero. For backwards compatibility\\
      MJD        & D (1)     & Modified Julian Day\\
      INT\_TIME  & D (1)     & Integration time (s)\\
      VIS2DATA   & D (NWAVE) & Squared Visibility\\
      VIS2ERR    & D (NWAVE) & Error in Squared Visibility\\
      CORRINDX\_VIS2DATA & J (1) & (optional) Index into correlation matrix for 1st VIS2DATA element\\
      UCOORD     & D (1)     & U coordinate of the data (m)\\
      VCOORD     & D (1)     & V coordinate of the data (m)\\
      STA\_INDEX & I (2)     & Station numbers contributing to the data\\
      FLAG       & L (NWAVE) & Flag\B\\
\hline\hline 
\end{tabular}
\end{table*}
\subsubsection*{Cross-referencing}
Cross-referencing between FITS tables is not a generic feature of the
FITS format, but is necessary for correct description of the
interferometric data. When writing OIFITS tables, particular care must
be taken to maintain the cross-references. Specifically, each data
table must refer:
\begin{itemize}
\item to a particular OI\_WAVELENGTH table describing the wavelength
  channels for the measurements, via the \texttt{INSNAME} keyword; and
\item to an OI\_ARRAY table, via the \texttt{ARRNAME} keyword; and
\item optionally, to the OI\_INSPOL table containing the same
  \texttt{INSNAME}; and
\item optionally, to an OI\_CORR table via the \texttt{CORRNAME} keyword.
\end{itemize}

There is a second level of cross-referencing possible, between table
elements.  Indeed, interferometric observables stored in an OIFITS file are
physically associated:
a phase closure in \texttt{T3PHI}, for example, can be the argument
of the closure of three complex values whose squared norms are stored
in three rows of the \texttt{VIS2DATA} column. Usually, the
\texttt{MJD} column contains the same timestamp for all the
simultaneous -- therefore, related -- observables, and suffices to
link them. It should be stressed however that, according to the
definition of \texttt{MJD} above, timestamps
for two related observables can differ: values in an OIFITS being
averaged and calibrated, the number of raw data used in the
average, or the calibration process, may change the mean UTC time of
the measurement even between related observables.

Data producers that want to add a more direct link than \texttt{MJD}
between related quantities in different tables can add a non-standard
column in the relevant tables, such as a \texttt{NS\_DATAINDEX}
column of \texttt{TFORMn} value '\texttt{J}', containing integer hash
numbers.
\subsubsection*{Flag}
If a value in this vector is true, the corresponding datum must be
ignored in all analysis.
\subsection{OI\_VIS2 (revision~2)}
\label{sect:vis2}
Keywords for the OI\_VIS2 table are listed in Table~\ref{tbl:oivis2}. 
\begin{table*}[t]
\caption{OI\_VIS (revision~$2$)}
\label{tbl:oivis}
\begin{tabular}[c]{l c l}
\hline\hline
\T \hfill Label \hfill &\hfill Data Type\hfill&\hfill Description\hfill \B\\ 
\hline
\multicolumn{3}{c}{\T Keywords \B}\\
\hline
\T 
      OI\_REVN  & \texttt{2} & Revision number:~$2$\\
      DATE-OBS  & \%s & UTC start date of observations\\
      ARRNAME   & \%s & Identifies corresponding OI\_ARRAY\\
      INSNAME   & \%s & Identifies corresponding OI\_WAVELENGTH table\\
      CORRNAME  & \%s & (optional$^{\dag}$) Identifies corresponding OI\_CORR table\\
      AMPTYP    & \%s & (optional$^{\dag}$) ``absolute'',``differential'',``correlated~flux''\\
      PHITYP    & \%s & (optional$^{\dag}$) ``absolute'', ``differential'' \\
      AMPORDER & \%d & (optional$^{\dag}$) Polynomial fit order for differential chromatic amplitudes\\
      PHIORDER & \%d & (optional$^{\dag}$) Polynomial fit order for differential chromatic phases\B\\
\hline
\multicolumn{3}{c}{\T Column Headings (one row for each measurement)\B}\\
\hline
\T 
      TARGET\_ID & I (1) & Target number as index into OI\_TARGET table\\
      TIME       & D (1) & Zero. For backwards compatibility\\
      MJD        & D (1) & Modified Julian Day\\
      INT\_TIME  & D (1) & Integration time (s)\\
      VISAMP     & D (NWAVE) & Visibility amplitude\\
      VISAMPERR  & D (NWAVE) & Error in visibility amplitude\\
      CORRINDX\_VISAMP & J (1) & (optional) Index into correlation matrix for 1st VISAMP element\\
      VISPHI     & D (NWAVE) & Visibility phase in degrees\\
      VISPHIERR  & D (NWAVE) & Error in visibility phase in degrees\\
      CORRINDX\_VISPHI & J (1) & (optional) Index into correlation matrix for 1st VISPHI element\\
      VISREFMAP & L (NWAVE,NWAVE) & (optional$^{\dag}$) Matrix of indexes for establishing the reference channels (see text).\\
      RVIS & D (NWAVE) & (optional) Complex coherent flux (Real) in units of TUNITn\\ 
      RVISERR  & D (NWAVE) & (optional) Error on RVIS \\
      CORRINDX\_RVIS & J (1) & (optional) Index into correlation matrix for 1st RVIS element\\
      IVIS & D (NWAVE) & (optional) Complex coherent flux (Imaginary) in units of TUNITn\\ 
      IVISERR  & D (NWAVE) & (optional) Error on IVIS \\
      CORRINDX\_IVIS & J (1) & (optional) Index into correlation matrix for 1st IVIS element\\
      UCOORD     & D (1) & U coordinate of the data (m)\\
      VCOORD     & D (1) & V coordinate of the data (m)\\
      STA\_INDEX & I (2) & Station numbers contributing to the data\\
      FLAG       & L (NWAVE) & Flag\B\\
\hline\hline 
\end{tabular}
{\par\T\textbf{Note.} $\dag$~The conditional use of these keywords is discussed in Sect.~\ref{sect:vis}.}
\end{table*}
\subsubsection*{Data arrays}
\texttt{NWAVE} is the number of spectral channels, given by the
\texttt{NAXIS2} keyword of the relevant OI\_WAVELENGTH table.
\subsubsection*{UV  coordinates}
\texttt{UCOORD}, \texttt{VCOORD} give the $(u,v)$ coordinates in meters
of the point in the UV plane associated with the vector of
visibilities. The data points may be averages over some region of the
UV plane, but the current version of the standard says nothing about
the averaging process. 
\subsection{OI\_VIS (revision~2)}
\label{sect:vis}

Keywords for the OI\_VIS table are listed in Table~\ref{tbl:oivis}.
This table has been used in the past to store the amplitude and phase
of visibilities obtained through phase-preserving coherent
integration. This revision intends to enable storage of complex
coherent fluxes, visibilities and differential chromatic visibilities,
uniquely identified so they can be modeled without ambiguity.
\subsubsection*{New keywords and columns}
There are four new keywords in the header, \texttt{AMPTYP},
\texttt{PHITYP}, \texttt{AMPORDER}, and
\texttt{PHIORDER}. 

\texttt{AMPTYP} defines the type of data stored in
\texttt{VISAMP}: \texttt{absolute}, \texttt{differential} or
\texttt{correlated flux}. In the last case the usual
\texttt{TUNIT}n FITS keyword must be used to specify the flux unit
(such as \texttt{photon}, \texttt{count}, \texttt{Jy},
\texttt{adu}, see footnote\,\ref{ivoaspectraldm}). 

\texttt{PHITYP} is one of \texttt{absolute} or
\texttt{differential}. The default for \texttt{AMPTYP} and
\texttt{PHITYP}, if not present, is \texttt{absolute}.

If \texttt{AMPTYP} or \texttt{PHITYP} is \texttt{differential},
the newly-defined data column \texttt{VISREFMAP} must be present since
this column contains the channel number(s) whose combination
defines the reference channel used in computing the
\texttt{VISAMP} and \texttt{VISPHI} values. The $i^{th}$ row of the
2-D matrix \texttt{VISREFMAP} specifies the reference channels for
channel $i$. (In FITS the first dimension corresponds to the
  most rapidly varying index, also known as column-major
  ordering. This internal ordering is usually hidden by the low-level
  FITS libraries such as \texttt{cfitsio}.)

The combination of the channels in \texttt{VISREFMAP} defaults to a
simple average, unless a polynomial fit of some order is defined
through \texttt{AMPORDER} or \texttt{PHIORDER}.

If both \texttt{VISAMP} and \texttt{VISPHI} are differential,
\texttt{VISREFMAP} shall apply to both. \texttt{VISREFMAP} must be
present if either \texttt{VISTYP} or \texttt{PHITYP} is
\texttt{differential}. If \texttt{VISREFMAP} is absent, neither
can be differential.

OI\_VIS revision~2 implements new optional columns
\texttt{RVIS} and \texttt{IVIS} allowing the storage of complex
visibilities or complex fluxes in (Real, Imaginary) representation
rather than amplitude and phase.
\texttt{RVIS}, \texttt{IVIS} and their associated errors
\texttt{RVISERR} and \texttt{IVISERR} may be omitted. Their unit is given by the
standard table header \texttt{TUNIT}n keywords.  Covariances between
elements of the \texttt{RVIS} and \texttt{IVIS} vectors can be stored
using the general OI\_CORR sparse correlation matrix mechanism via the
\texttt{CORRINDX\_RVIS} and \texttt{CORRINDX\_IVIS} columns.
\begin{table*}[htbp]
\caption{OI\_T3 (revision~$2$)}
\label{tbl:oit3}
\begin{tabular}[c]{l c l}
\hline\hline
\T \hfill Label \hfill &\hfill Data Type\hfill&\hfill Description\hfill \B\\ 
\hline
\multicolumn{3}{c}{\T Keywords \B}\\
\hline
\T 
      OI\_REVN  & \texttt{2} & Revision number:~$2$\\
      DATE-OBS  & \%s & UTC start date of observations\\
      ARRNAME   & \%s & Identifies corresponding OI\_ARRAY\\
      INSNAME   & \%s & Identifies corresponding OI\_WAVELENGTH table\\
      CORRNAME  & \%s & (optional) Identifies corresponding OI\_CORR table\B\\
\hline
\multicolumn{3}{c}{\T Column Headings (one row for each measurement)\B}\\
\hline
\T 
      TARGET\_ID & I (1) & Target number as index into OI\_TARGET table\\
      TIME       & D (1) & Zero. For backwards compatibility\\
      MJD        & D (1) & Modified Julian Day\\
      INT\_TIME  & D (1) & Integration time (s)\\
      T3AMP      & D (NWAVE) & Triple Product Amplitude\\
      T3AMPERR   & D (NWAVE) & Error in Triple Product Amplitude\\
      CORRINDX\_T3AMP & J (1) & (optional) Index into correlation matrix for 1st T3AMP element\\
      T3PHI      & D (NWAVE) & Triple Product Phase in degrees\\
      T3PHIERR   & D (NWAVE) & Error in Triple Product Phase in degrees\\
      CORRINDX\_T3PHI & J (1) & (optional) Index into correlation matrix for 1st T3PHI element\\
      U1COORD    & D (1) & U coordinate of baseline AB of the triangle (m)\\
      V1COORD    & D (1) & V coordinate of baseline AB of the triangle (m)\\
      U2COORD    & D (1) & U coordinate of baseline BC of the triangle (m)\\
      V2COORD    & D (1) & V coordinate of baseline BC of the triangle (m)\\
      STA\_INDEX & I (3) & Station numbers contributing to the data\\
      FLAG       & L (NWAVE) & Flag\B\\
\hline\hline 
\end{tabular}
\end{table*}
\subsubsection*{Data arrays}
As in revision\,$1$,
\texttt{FLAG} applies to \texttt{VISAMP}, \texttt{VISAMPERR},
\texttt{VISPHI} and \texttt{VISPHIERR}.  \texttt{VISAMP} and
\texttt{VISPHI}, as well as their associated errors \texttt{VISAMPERR}
and \texttt{VISPHIERR}, can be \texttt{NULL} in the cases where the
flag mechanism is not applicable, for example, if the instrument
computes only \texttt{VISPHI}, or if flagging some incorrect
\texttt{VISAMP} values would flag otherwise good \texttt{VISPHI}
values.  Similarly, if some values of \texttt{RVIS} or \texttt{IVIS}
or \texttt{RVISERR} or \texttt{IVISERR} are meaningless, \texttt{NULL}
values are to be used instead.

\texttt{NWAVE} is the number of spectral channels, given by the
\texttt{NAXIS2} keyword of the relevant OI\_WAVELENGTH table.
\subsubsection*{UV  coordinates}
\texttt{UCOORD}, \texttt{VCOORD} give the $(u,v)$ coordinates in meters
of the point in the UV plane associated with the vector of
visibilities -- see Sect.~\ref{sect:defn} for details. The data points
may be averages over some region of the UV plane, but the current
version of the standard says nothing about the averaging process.
\subsection{OI\_T3 (revision~2)}
\label{sect:t3}
Keywords for the OI\_T3 table are listed in Table~\ref{tbl:oit3}. 
\subsubsection*{Data arrays}
\texttt{FLAG} marks both the corresponding bispectrum amplitude
(\texttt{T3AMP}) and phase (\texttt{T3PHI}) as invalid. To mark only
the amplitude or phase as invalid, the relevant data values must be
replaced by \texttt{NULL} (with \texttt{FLAG} remaining as false). If the
dataset does not provide triple product amplitudes at all,
\texttt{T3AMP} and \texttt{T3AMPERR} must contain only \texttt{NULL}
values.

\texttt{NWAVE} is the number of spectral channels, given by the
\texttt{NAXIS2} keyword of the relevant OI\_WAVELENGTH table.
\subsubsection*{Triple product UV coordinates}
The \texttt{U1COORD}, \texttt{V1COORD}, \texttt{U2COORD}, and
\texttt{V2COORD} columns contain the $(u,v)$ coordinates of the
bispectrum point in meters. \texttt{U3COORD} and
\texttt{V3COORD} are not given
since $u_1+u_2+u_3=0\,,\,v_1+v_2+v_3=0$. The corresponding data
points may be averages in (bi-) spatial frequency space, but this
version of the standard does not attempt to describe the averaging
process.
\section{Optional Tables}
\subsection{OI\_FLUX (revision~1)}
\label{sect:flux}
\begin{table*}[t]
\caption{OI\_FLUX (revision~$1$)}
\label{tbl:oiflux}
\begin{tabular}[c]{l c l}
\hline\hline
\T \hfill Label \hfill &\hfill Data Type\hfill&\hfill Description\hfill \B\\ 
\hline
\multicolumn{3}{c}{\T Keywords \B}\\
\hline
\T 
OI\_REVN  & \texttt{1} & Revision number: 1\\
DATE-OBS  & \%s & UTC start date of observations\\
INSNAME   & \%s & Identifies corresponding OI\_WAVELENGTH table\\
ARRNAME   & \%s & (optional$^{\dag}$) Identifies corresponding OI\_ARRAY table\\
CORRNAME  & \%s & (optional) Identifies corresponding OI\_CORR table\\
FOV	& \%f & (optional$^{\dag}$) Area on sky over which flux is integrated (arc\,sec)\\
FOVTYPE	   & \%s & (optional$^{\dag}$) Model for FOV: ``FWHM'' or ``RADIUS''  \\
CALSTAT & \%c & ``C'': Spectrum is calibrated, ``U'': uncalibrated\B\\
\hline
\multicolumn{3}{c}{\T Column Headings (one row for each source)\B}\\
\hline
\T 
TARGET\_ID  & I (1) & Target number as index into OI\_TARGET table\\
MJD         & D (1) & Modified Julian Day\\
INT\_TIME   & D (1) & Integration time (s)\\
FLUXDATA    & D (NWAVE) & Flux in units of TUNITn\\
FLUXERR     & D (NWAVE) & Corresponding flux error\\
CORRINDX\_FLUXDATA & J (1) & (optional) Index into correlation matrix for 1st FLUXDATA element\\
STA\_INDEX  & I (1) & (optional$^{\dag}$) Station number contributing to the data\B\\
      FLAG       & L (NWAVE) & Flag\B\\
\hline\hline 
\end{tabular}
{\par\T \textbf{Note.} $\dag$ The conditional use of these keywords is discussed in Sect.~\ref{sect:flux}.}
\end{table*}
OIFITS version\,$2$ introduces this optional table intended as a
container for raw or calibrated flux measurements. Corresponding
wavelengths are referenced with the \texttt{INSNAME} keyword. The
content of the table depends on the single-character keyword header
\texttt{CALSTAT}.  Keywords for the OI\_FLUX table are listed in
Table~\ref{tbl:oiflux}. \texttt{FLUXDATA} and \texttt{FLUXERR} are in
units of their corresponding \texttt{TUNIT}, which must be
present\footnote{\label{ivoaspectraldm}See Appendix B.1 of the
  IVOA Spectral Data Model document available at
  \url{http://www.ivoa.net/documents/SpectralDM} for examples. 
  A FITS empty string (a minimum of eight whitespaces) is a valid value
  for a \texttt{TUNIT} keyword.}.

If \texttt{CALSTAT} is \texttt{C} (Calibrated), the fluxes listed
are flux-calibrated spectra obtained with the interferometric
instrument itself, if it provides such a measurement, or with another
instrument. In both cases this spectrum is useful for modeling sources
with physical emission components where it provides the constraint of
the total flux. The spectrum is the spectrum of the object and does
not depend on the telescope used: the table shall not contain
\texttt{ARRNAME} and \texttt{STA\_INDEX}.  

If \texttt{CALSTAT} is \texttt{U} (Uncalibrated), \texttt{ARRNAME}
and the \texttt{STA\_INDEX} column must be present, and the flux
stored in \texttt{FLUXDATA} is the flux measured on the telescope of
array \texttt{ARRNAME} indicated by \texttt{STA\_INDEX}. In this case,
the table shall not contain \texttt{FOV} and \texttt{FOVTYPE}. (The
field of view is given in the OI\_WAVELENGTH table referenced by
\texttt{INSNAME}.)
\subsubsection*{Data arrays}
\texttt{NWAVE} is the number of distinct channels, given by the
\texttt{NAXIS2} keyword of the relevant OI\_WAVELENGTH table.
\begin{table}[b]
  \caption{OI\_CORR (revision~$1$)}
  \label{tab:CORR}
    \begin{tabular}{lcl}
      \hline
      \hline
     \T Label & Data Type & Description \B \\
      \hline
      \multicolumn{3}{c}{\T Keywords \B} \\
      \hline
      \T OI\_REVN & \texttt{1} & Revision number: 1\\
      CORRNAME & \%s & Name of correlated data set$^{(1)}$\\
      NDATA  & \%d & Size of the correlation matrix\B \\
      \hline
      \multicolumn{3}{c}{\T Column Headings, one row for each non-zero element} \B\\
      \hline
      \T IINDX & J(1) & First index of correlation matrix element\\
      JINDX & J(1) & Second index of correlation matrix element\\
      CORR  & D(1) & Matrix element (IINDX,JINDX) \B\\
      \hline
      \hline
    \end{tabular}
{\par\T\small \textbf{Notes.} $^{(1)}$~this name cross-references one or more tables in the file.}
\end{table}
\subsection{OI\_CORR (revision~1) }
There is a need to describe correlations between any kinds of OIFITS
observables, over a limited timespan (hours at most) determined by the
calibration procedures. For an example where such correlations are
derived, see \cite{2003A&A...400.1173P}.

The following addition to the standard allows the definition of a
set of correlated real-valued data $\{x_i\}$ that may span
multiple OIFITS tables, of any type. Data may be correlated in
wavelength, baseline, or time. Correlations between different kinds of
observable can also be described. An OIFITS file may contain multiple
such sets, to accommodate merging of multiple data sets. The
correlations are described by a sparse dimensionless correlation
  matrix
\begin{equation}
C_{ij} = \frac{\sigma^2_{ij}}{\sqrt{\sigma^2_i \sigma^2_j}} ,
\end{equation}
which is stored in a new OI\_CORR table (Table~\ref{tab:CORR}). The
covariance matrix can straightforwardly be obtained from the
correlation matrix using the standard deviations ${\sqrt{\sigma^2_i}}$
from the data tables, such as \texttt{VIS2ERR} in OI\_VIS2 which is the
standard deviation of the squared visibility.

Because the number of potentially correlated data is relatively large
-- of order $10^4$;
a six-element interferometer measuring squared visibilities, triple
amplitudes and closure phases in~100 spectral channels for two hours
with an observing cadence of 20 minutes would generate a number of
data \texttt{NDATA}=33,000 -- and at the same time many of these
correlations are not significant, we have adopted a sparse
representation of the correlation matrix. Only non-zero elements of
the \texttt{NDATA$\times$NDATA}
matrix are stored as triplets $\{i, j, C_{ij}\}$.
Since $C_{ij} = C_{ji}$
and $C_{ii} = 1$,
we adopt the convention of only storing elements with $j > i$.
The indices $i$
and $j$
are stored as signed 32-bit integers (J type code in
\texttt{TFORM}n). (The indices are all positive but unsigned integers
can only be represented in FITS by specifying an offset using the
\texttt{BZERO} keyword.)
\begin{table*}[htbp]
\caption{OI\_INSPOL (revision~$1$)}
\label{tbl:inspolar}
\begin{tabular}[c]{l c  l}
\hline\hline
\T \hfill Label \hfill &\hfill Data Type\hfill&\hfill Description\hfill \B\\ 
\hline
\multicolumn{3}{c}{\T Keywords \B}\\
\hline
\T 
OI\_REVN  & \texttt{1} & Revision number: 1\\
NPOL      & \%d & Number of Polarisation Types in this table\\
ARRNAME   & \%s & Identifies corresponding OI\_ARRAY\\
ORIENT & \%s & Orientation of the Jones Matrix, could be ``NORTH'' (for on-sky orientation), or ``LABORATORY''\\
MODEL & \%s & A string keyword that describe the way the Jones matrix is estimated\B\\
\hline
\multicolumn{3}{c}{\T Column Headings, one row for each source\B}\\
\hline
\T 
TARGET\_ID & I (1) & Target number as index into OI\_TARGET table\\
INSNAME    & A     & INSNAME of this polarisation \\
MJD\_OBS    & D (1) & Modified Julian Day, start of time lapse\\
MJD\_END    & D (1) & Modified Julian Day, end of time lapse\\
JXX        & C (NWAVE) & Complex Jones Matrix component along X axis.\\
JYY        & C (NWAVE) & Complex Jones Matrix component along Y axis\\
JXY        & C (NWAVE) & Complex Jones Matrix component between X and Y axis\\
JYX        & C (NWAVE) & Complex Jones Matrix component between Y and X axis\\
STA\_INDEX & I (1) & Station number for the above matrices\B\\
\hline\hline 
\end{tabular}
\end{table*}

Each of the data tables in a correlated data set shall have a
keyword \texttt{CORRNAME} that is used to look up the corresponding
OI\_CORR table. The correspondence between the position of a value in
a data table and its index into the correlation matrix is given by the
following new columns: \texttt{CORRINDX\_VISAMP},
\texttt{CORRINDX\_VISPHI}, \texttt{CORRINDX\_RVIS} and
\texttt{CORRINDX\_IVIS} in OI\_VIS, \texttt{CORRINDX\_VIS2DATA} in
OI\_VIS2, \texttt{CORRINDX\_T3AMP} and \texttt{CORRINDX\_T3PHI} in
OI\_T3 and \texttt{CORRINDX\_FLUXDATA} in OI\_FLUX. These columns
are all of data type J(1) and contain values $\geq1$. If all of the data in a table are
uncorrelated, the \texttt{CORRNAME} keyword and all of the
\texttt{CORRINDX\_*} columns shall be omitted. If some of the table
data are correlated, correlation matrix indices shall be specified for
all of the data in the table, but the zero-valued correlations shall
be omitted from the OI\_CORR table.

Taking the example of an OI\_VIS2 table whith correlated values, the
(single) \texttt{CORRINDX\_VIS2DATA} value in a particular row in
OI\_VIS2 gives the index into the correlation matrix for the first
element of the \texttt{VIS2DATA} vector of this row, which contains \texttt{NWAVE}
values for the defined spectral channels. The index for the $j$th
value of the data vector is given by
$(\verb+CORRINDX_VIS2DATA+ + j - 1)$.
In other words, an element in the correlation matrix with indices
$(i,j)$
gives the value of the correlation between the \texttt{VIS2DATA}
measurement in the row of the referencing OI\_VIS2 with the largest
\texttt{CORRINDX\_VIS2DATA} value $i_r$
less than or equal to $i$
and channel index $(i-i_r+1)$
and the measurement in OI\_VIS2 obtained in the same way using index
$j$
instead of $i$.
Therefore, the indices implied by the \texttt{CORRINDX} values shall
be unique within the correlated data set.  The other \texttt{CORRINDX}
columns give the correlation matrix index for the first element of the
similarly-named, real-valued data column. An example of use is
described in Appendix~\ref{app:example}.
\subsection{OI\_INSPOL (revision~1)}
\label{sect:inspolar}
In general, each input beam contributing to the correlated flux on a
baseline is polarized to some extent, either by design or by the
unavoidable effects of partially-polarizing elements in the optical
train of the interferometer. Although polaro-interferometric
observations in the optical and infrared range are possible, there is no good
example yet of true polaro-interferometric observations which provided
a measurement of the source's Stokes parameters.  It is thus out of
the scope of this document to standardize polaro-interferometric
calibrated data.  However, we feel compelled to provide a way to note
in the OIFITS format that the measurements were subject to
instrumental polarisation, and, in the case of simultaneous or related
measurements in various states of polarisation, provide an internal
link between the classical observables of OIFITS and the
instrumental polarisation state in which they were obtained. This new
table is purely instrumental and named OI\_INSPOL.

In order to ensure backwards compatibility with current OIFITS
readers, we identify each functional polarisation state by a different
\texttt{INSNAME}. For example, data taken through a Wollaston prism
(two orthogonal linear polarisations) provides two statistically
independent series of interferograms. Each serie will produce OI
observables, perfectly valid independently of the other. It is just a
matter of bookkeeping to note that one serie was taken with
horizontal polarisation through the Wollaston and the other
through the vertical polarisation.  The new OI\_INSPOL table is
the central hub that links all the \texttt{INSNAME}s (all the
polarized measurements).  Keywords for the OI\_INSPOL table are listed
in Table~\ref{tbl:inspolar}.

The OI\_INSPOL table uses the Jones matrix representation as a
beam-based description of how the optical system operates on the
source's light. This formalism does not permit the description of the
depolarizing effects that could occur between the telescope(s) and the
recording of the interferogram. However, being an operator on the
electrical field of the light, it allows to describe the effects on
the complex coherence betwen two beams.

\begin{table}[b]
\caption{Basic Jones Matrix Examples. See text for the sign of imaginary parts.}
\label{tab:ExamplePolar}
{\footnotesize
\begin{tabular}{|c|c c c c|}
\hline
\T Polarizer & JXX & JXY & JYX & JYY \B\\
\hline
\T Linear, Horizontal& $(1,0)$ & $(0,0)$ & $(0,0)$ & $(0,0)$\\
Linear, Vertical &  $(0,0)$ & $(0,0)$ & $(0,0)$ & $(1,0)$\\
Linear, $\pm45\deg$&  $(0.5,0)$& $(\pm0.5,0)$& $(\pm0.5,0)$& $(0.5,0)$\\
Circular, Right&  $(0.5,0)$& $(0,0.5)$& $(0,-0.5)$& $(0.5,0)$\\
Circular, Left&  $(0.5,0)$& $(0,-0.5)$& $(0,0.5)$& $(0.5,0)\B$\\
\hline
\end{tabular}
}
\end{table}
The OI\_INSPOL table stores a Jones matrix for each wavelength, time
interval, and polarisation state as cross-referenced via the
\texttt{INSNAME} column.  The four complex elements of the Jones matrix
are stored in the columns \texttt{JXX}, \texttt{JYY}, \texttt{JXY} and
\texttt{JYX}. These values are beam-based, and can be in principle
measured or modeled as a function of time. In the simplest case, a
polarizer in front of the camera would record only one polarisation,
or a Wollaston in the same place would permit recording of two
orthogonally polarized interferograms. Then OI\_INSPOL would serve
only to describe the \texttt{ORIENT}ation of the polarisation state of
the \texttt{INSNAME} (or \texttt{INSNAME}s) interferometric data. More
complicated cases, in particular a time-varying instrumental
polarisation due to rotation of a polarizing element, are
possible. The \texttt{MJD\_OBS} and \texttt{MJD\_END} values can
describe any variation of the complex Jones matrices independently of
the time-tagging of data contained the other
tables. Table~\ref{tab:ExamplePolar} shows a few examples of Jones
matrices for simple polarizers. The sign of the imaginary parts must
be given in the theoretical framework where the wave function used in
propagation equations of light reads $\exp(k{z}-i\omega{t})$
\cite[p.~10]{PBLS}.

OI\_INSPOL is an optional table: if OI\_INSPOL is present, it must
encompass at least all the time entries and baselines present in all
the other tables referenced by all the values stored in its
\texttt{INSNAME} column. OI\_INSPOL is also experimental: OIFITS
readers or writers should make provision for the possible augmentation of this
table in future releases.
\begin{table*}[tp]
  \caption{This table gives an example of a set of data comprising
    two OI\_T3 tables obtained at two different times, for four channels
    on three baselines, and the corresponding \texttt{VIS2DATA} measurements,
    also in two OI\_VIS2 tables. There are three cases examined in the text
    (Appendix~\ref{app:example}), which lead to four different OI\_CORR tables.
    (The corresponding \texttt{CORRNAME} is given for each.)
  }
\label{tab:ExampleCorr}
\begin{tabular}{|c|c|c|c|cccc|}
\hline
\T & & &  & $\lambda_1$&$\lambda_2$&$\lambda_3$&$\lambda_4$\B\\
\cline{5-8}
\T MJD & CORRNAME & CORRINDX & STA\_INDEX & \multicolumn{4}{c|}{T3}\B\\
\hline
\multicolumn{8}{c}{\T Table OI\_T3 \#1 \B}\\
\hline
\T 1.0 & \X{``T1''}{``V\&T''} & \X{1}{1} & 1 2 3 & $\mathrm{T3}^{mjd=1.0}_{\lambda_1}$ & $\mathrm{T3}^{mjd=1.0}_{\lambda_2}$ & $\mathrm{T3}^{mjd=1.0}_{\lambda_3}$ & $\mathrm{T3}^{mjd=1.0}_{\lambda_4}$ \B\\
\hline
\multicolumn{8}{c}{\T Table OI\_T3 \#2 \B}\\
\hline
\T 2.0  & \X{``T2''}{``V\&T''} &\X{1}{5}  & 1 2 3 & $\mathrm{T3}^{mjd=2.0}_{\lambda_1}$ & $\mathrm{T3}^{mjd=2.0}_{\lambda_2}$ & $\mathrm{T3}^{mjd=2.0}_{\lambda_3}$ & $\mathrm{T3}^{mjd=2.0}_{\lambda_4}$ \B\\
\hline
\multicolumn{8}{c}{\T $\vdots$ \B}\\
\hline
\T & & &  & $\lambda_1$&$\lambda_2$&$\lambda_3$&$\lambda_4$\B\\
\cline{5-8}
\T     MJD & CORRNAME & CORRINDX & {STA\_INDEX} & \multicolumn{4}{c|}{VIS2DATA} \B\\
\hline
\multicolumn{8}{c}{\T Table OI\_VIS2 \#1 \B}\\
\hline
\T 1.0  & \X{``V''}{``V\&T''} & \X{1}{9} & 1 2 &  $\mathrm{V2}_{12,\lambda_1}^{mjd=1.0}$ & $\mathrm{V2}_{12,\lambda_2}^{mjd=1.0}$ & $\mathrm{V2}_{12,\lambda_3}^{mjd=1.0}$ & $\mathrm{V2}_{12,\lambda_4}^{mjd=1.0}$ \\
\hline
1.0  & \X{``V''}{``V\&T''} & \X{5}{13} & 2 3 &  $\mathrm{V2}_{23,\lambda_1}^{mjd=1.0}$ & $\mathrm{V2}_{23,\lambda_2}^{mjd=1.0}$ & $\mathrm{V2}_{23,\lambda_3}^{mjd=1.0}$ & $\mathrm{V2}_{23,\lambda_4}^{mjd=1.0}$ \\
\hline
1.0  & \X{``V''}{``V\&T''} & \X{9}{17} & 1 3 &  $\mathrm{V2}_{13,\lambda_1}^{mjd=1.0}$ & $\mathrm{V2}_{13,\lambda_2}^{mjd=1.0}$ & $\mathrm{V2}_{13,\lambda_3}^{mjd=1.0}$ & $\mathrm{V2}_{13,\lambda_4}^{mjd=1.0}$ \B\\
\hline
\multicolumn{8}{c}{\T Table OI\_VIS2 \#2\B}\\
\hline
\T 2.0 & \X{``V''}{``V\&T''} & \X{13}{21} & 1 2&  $\mathrm{V2}_{12,\lambda_1}^{mjd=2.0}$ & $\mathrm{V2}_{12,\lambda_2}^{mjd=2.0}$ & $\mathrm{V2}_{12,\lambda_3}^{mjd=2.0}$ & $\mathrm{V2}_{12,\lambda_4}^{mjd=2.0}$ \\
\hline
2.0 & \X{``V''}{``V\&T''} & \X{17}{25} & 2 3 &    $\mathrm{V2}_{23,\lambda_1}^{mjd=2.0}$ & $\mathrm{V2}_{23,\lambda_2}^{mjd=2.0}$ & $\mathrm{V2}_{23,\lambda_3}^{mjd=2.0}$ & $\mathrm{V2}_{23,\lambda_4}^{mjd=2.0}$ \\
\hline
2.0 & \X{``V''}{``V\&T''} & \X{21}{29} & 1 3 &    $\mathrm{V2}_{13,\lambda_1}^{mjd=2.0}$ & $\mathrm{V2}_{13,\lambda_2}^{mjd=2.0}$ & $\mathrm{V2}_{13,\lambda_3}^{mjd=2.0}$ & $\mathrm{V2}_{13,\lambda_4}^{mjd=2.0}$  \B\\
\hline
\end{tabular}
\end{table*}
\subsection{Guidelines for other tables}
It may be useful to incorporate other tables in an OIFITS file. For
example, there might be one that contains instrument specific
information, such as the backend configuration. Another optional table
could contain information relevant to astrometry.  The only
requirement imposed by the format is that \texttt{EXTNAME}s of additional
tables must not begin with ``OI\_''.
\section{Summary}
Drawing on experience from a dozen years of community use, we have
revised the OIFITS data format to:
\begin{itemize}
\item address the needs of the new interferometric instruments that
  will routinely produce images, adding ancillary data and tables
  useful for image reconstruction purposes;
\item provide a rigorous description of the interferometric
  observables and how they are computed, especially in the case of
  differential observables;
\item provide a rigorous description of measurement errors and their
  correlation;
\item tentatively take into account simple polarisation properties of
  the instrument;
\item pave the way to a more formal description of the optical
  interferometry Data Model, as developed for radio interferometry for
  example.
\item add bookkeeping keywords necessary to feed the
  emerging databases of optical interferometry measurements.
\end{itemize}
This has been done while guaranteeing backward compatibility with
version\,$1$, ensuring continuous use of older software without
changes. There are already a few libraries available for reading and
writing this format, and we expect that it will soon be used by the
end data products of the PIONIER, GRAVITY and MATISSE instruments, and
receive wide support from the optical interferometry community.
%
%

\begin{appendix}
\section{On the use of Correlation Matrices in OIFITS}
\label{app:example}
We describe here a concrete example of the use of OI\_CORR. Consider a
dataset comprising two OI\_VIS2 tables and two OI\_T3 tables, as shown
in Table~\ref{tab:ExampleCorr}. Each table contains observations made
simultaneously on three baselines in four wavelength
channels. Depending on how we want to express the correlations between
some of the observables within these tables, we will use different
OI\_CORR tables (with unique \texttt{CORRNAME}s).

If we need to express the spectral correlations of the T3AMP
measurements only, we would write a $4\times4$ full correlation
matrix for each OI\_T3 table. These two matrices can be written as two
sparse matrices in two separate OI\_CORR tables, ``T1'' and ``T2''
(one per OI\_T3). The value of \texttt{CORRINDX\_T3AMP} would be~$1$ for
both OI\_T3 tables as shown in Table~\ref{tab:ExampleCorr} in the
lower diagonal cells.

If we need to express the time and baseline correlations between the
\texttt{VIS2DATA} measurements only, we have
$2 \times 3 \times 4 = 24$ measurements in the two OI\_VIS2 tables
($12$ per table), requiring a $24\times24$ correlation matrix.  The
correlations between these measurements would be expressed as a single
OI\_CORR table (``V'') with $\verb+NDATA+ = 24$. In this example, the
\texttt{CORRINDX\_VIS2DATA} values for each row in OI\_VIS2 are also
given in the lower diagonal cells of Table~\ref{tab:ExampleCorr}.

As \texttt{T3AMP} measurements share baselines with \texttt{VIS2DATA}
measurements, we could give the correlations between \texttt{T3AMP}
and \texttt{VIS2} measurements in a single OI\_CORR table. In
this case, non-zero correlations would be contained in a matrix of
$\verb+NDATA+ = 24 + 2\times4 = 32$ rows and columns, the indices
\texttt{CORRINDX\_VIS2DATA} and \texttt{CORRINDX\_T3AMP} are now given
in the upper diagonal cells of Table~\ref{tab:ExampleCorr} and all
data tables would reference the same OI\_CORR table (``V\&T'').
\end{appendix}
\begin{acknowledgements} 
  Initial development of the format was done under the auspices
  of the IAU Working Group on Optical and infrared Interferometry. Version\,$2$
  has been pursued by the (discontinued) IAU Commission~54 and JMMC,
  with the active participation of many community interferometrists,
  in particular: Fabien Baron, Philippe B\'{e}rio, Laurent Bourg\`{e}s,
  Leonard Burtscher, Alain Chelli, Pierre Cruzal\`{e}bes, Mike Ireland,
  Charleen Kemps, Brian Kloppenborg, Sylvestre Lacour, Vincent Lapeyrere,
  Jean-Baptiste le~Bouquin, Guillaume Mella, Florentin Millour, John
  Monnier, J\"{o}rg-Uwe Pott, Antony Schutz, Michel Tallon, Theo
  ten~Brummelaar, Eric Thi\'{e}baut.  We thank I.~Percheron, A.~Dobrzycki
  and our anonymous referee for many useful comments.

  This work has benefitted of the support of the European
  Community's Seventh Framework Programme (FP7/2013-2016) under Grant
  Agreement 312430 (OPTICON) and of the French ANR POLCA,
  ANR-10-BLAN-0511.
\end{acknowledgements}
%
%
%
\end{document}